%% file: main.tex
\documentclass[sigplan,twocolumn]{acmart}
\renewcommand\footnotetextcopyrightpermission[1]{}
\usepackage{array} 
\usepackage{tikz}
\usepackage{amsmath}

\usepackage[]{hyperref}
\usepackage{amsfonts}
\usepackage{url}
\usepackage{xspace}
\usepackage{subfig}
\usepackage{multirow}
\usepackage{makecell}
\usepackage[ruled, linesnumbered, vlined]{algorithm2e}
\usepackage{listings}

\usepackage[T1]{fontenc}

\usepackage{enumitem}
\usepackage{balance}

\begin{document}

\makeatletter
\fancypagestyle{standardpagestyle}{%
  \fancyhf{}%
  \renewcommand{\headrulewidth}{0pt}%
  \fancyhead[LO,LE]{\ACM@linecountL}%
  \fancyhead[RO,RE]{\ACM@linecountR}%
  \if@ACM@printfolios
    \fancyfoot[C]{\thepage}%
  \fi
}
\fancypagestyle{firstpagestyle}{%
  \fancyhf{}%
  \renewcommand{\headrulewidth}{0pt}%
  \fancyhead[LO,LE]{\ACM@linecountL}%
  \fancyhead[RO,RE]{\ACM@linecountR}%
  \if@ACM@printfolios
    \fancyfoot[C]{\thepage}%
  \fi
}
\pagestyle{standardpagestyle}
\makeatother

\title[]{\huge \sys{}: Efficient Text-to-Image Workflow Serving on a Serverless Platform}

\author{
    \rm{Xiaoxiao Jiang}$^{\dag *}$,
    \rm{Suyi Li}$^{\dag * \#}$,
    \rm{Sheng Yao}$^{\dag}$,
     \rm{Tianyu Feng}$^{\dag}$,
    \rm{Lingyun Yang}$^{\dag}$,
    \rm{Dapeng Nie},
    \rm{Haoran Yang},
    \rm{Wei Wang}$^{\dag}$
    \\
    $^{\dag}$Hong Kong University of Science and Technology
    \quad Alibaba Group
}

\input{macros.tex}
\input{contents/0_abstract}

\maketitle

\input{contents/1_introduction}
\input{contents/2_background}

\input{contents/3_overview}
\input{contents/4_design}
\input{contents/5_implementation}
\input{contents/6_evaluation}

\input{contents/7_related_works}
\input{contents/8_conclusion}

\bibliographystyle{ACM-Reference-Format}
\bibliography{main}

\input{contents/9_appendix}

\end{document}

%% file: macros.tex
\newcommand{\sys}{\textsc{ServerlessT2I}}

\newcommand{\PHB}[1]{\noindent\textbf{#1}\hspace{.5em}} 
\newcommand{\PHM}[1]{\vspace{.4em} \noindent\textbf{#1}\hspace{.5em}} 

\newcommand{\secref}[1]{\S\ref{#1}}
\newcommand{\figref}[1]{Fig.~\ref{#1}}
\newcommand{\tabref}[1]{Table~\ref{#1}}
\newcommand{\thmref}[1]{Theorem~\ref{#1}}
\newcommand{\prgref}[1]{Program~\ref{#1}}
\newcommand{\algref}[1]{Algorithm~\ref{#1}}
\newcommand{\eqnref}[1]{Equation~\ref{#1}}
\newcommand{\clmref}[1]{Claim~\ref{#1}}
\newcommand{\lemref}[1]{Lemma~\ref{#1}}
\newcommand{\ptyref}[1]{Property~\ref{#1}}

\newcommand{\eg}{{e.g.\@\xspace}}
\newcommand{\ie}{{i.e.\@\xspace}}
\newcommand{\etc}{
        \@ifnextchar{.}
        \textit{etc}
        \textit{etc.\@\xspace}
}

\newcommand{\term}{\textsf}
\newcommand{\code}{\texttt}
\newcommand{\ths}{\textsuperscript{th}}
\newcommand{\circledtext}[1]{\raisebox{.5pt}{\textcircled{\raisebox{-.9pt} {#1}}}}

\newcommand{\todo}[1]{\noindent\textcolor{red}{[TODO: #1]}}
\newcommand{\todowriting}[1]{\noindent\textcolor{red}{[Writing: #1]}}
\newcommand{\todofigure}[1]{\noindent\textcolor{red}{[Figure: #1]}}
\newcommand{\todoexp}[1]{\noindent\textcolor{red}{[Experiment: #1]}}
\newcommand{\suyi}[1]{\noindent\textcolor{violet}{#1}}
\newcommand{\wei}[1]{\textcolor{red}{#1}}
\newcommand{\xiaoxiao}[1]{\textcolor{cyan}{#1}}

\setlength{\abovecaptionskip}{3pt plus 1pt minus 1pt}
\setlength{\belowcaptionskip}{3pt plus 1pt minus 1pt}
\setlength{\abovedisplayskip}{3pt}
\setlength{\belowdisplayskip}{3pt}

\newcommand{\TeaCache}{\textsc{TeaCache}\xspace}
\newcommand{\Ideal}{\textsc{Ideal}\xspace}
\newcommand{\diffusers}{\textsc{Diffusers}\xspace}
\newcommand{\FISEdit}{\textsc{FISEdit}\xspace}
\newcommand{\vllm}{\textsc{vLLM-Omni}\xspace}
\newcommand{\noCache}{\textsc{NoCache}\xspace}
\newcommand{\wCache}{\textsc{WCache}\xspace}
\newcommand{\mCache}{\textsc{MCache}\xspace}

\definecolor{mygreen}{rgb}{0,0.6,0}
\definecolor{mygray}{rgb}{0.5,0.5,0.5}
\definecolor{mymauve}{rgb}{0.58,0,0.82}
\definecolor{myred}{rgb}{0.79,0.15,0.15}
\definecolor{myblue}{rgb}{0.1,0.1,0.8}
\lstdefinestyle{mypython}{
  language=Python,
  backgroundcolor=\color{white},   
  basicstyle=\scriptsize\ttfamily,        
  breakatwhitespace=false,         
  breaklines=true,                 
  captionpos=b,                    
  commentstyle=\color{mygreen},
  deletekeywords={...},            
  escapeinside={\%*}{*)},          
  extendedchars=true,              
  firstnumber=1,                
  frame=no,                    
  keepspaces=true,                 
  keywordstyle=\color{mymauve},       
  morekeywords={DIRECT, PERIODIC, IMMEDIATE, BY_TIME, EVERY_OBJ},            
  numbers=left,                    
  numbersep=5pt,                   
  numberstyle=\tiny\color{mygray}, 
  rulecolor=\color{black},         
  showspaces=false,                
  showstringspaces=false,          
  showtabs=false,                  
  stepnumber=1,                    
  stringstyle=\color{myred},     
  tabsize=4,                    
  title=\lstname                   
}

%% file: contents/0_abstract.tex
\begin{abstract}
Text-to-image (T2I) workflows are increasingly deployed on serverless platforms because users often compose customized workflows and invoke them intermittently.
Existing platforms typically deploy each workflow as an opaque GPU function, provisioning, placing, and scaling all constituent models in the workflow together. 
This monolithic design obscures workflow structure, inflates scaling overhead, forces users to manage low-level GPU coordination, and limits fine-grained fairness in multi-tenant clusters. 
In this paper, we present \sys{}, a serverless-native system that decomposes a T2I workflow into loosely coupled \emph{model functions} that can be independently managed and scheduled. 
By explicitly managing individual model execution, \sys{} enables per-model scaling, declarative workflow composition, transparent GPU-resident communication, and fairness-aware scheduling. 
To make this decomposition efficient, \sys{} harvests slack GPU memory left idle by compute-bound T2I inference to build a data plane that reduces model loading and data communication overheads. 
\sys{} further introduces a fair scheduler for multi-tenant serving.
Using production traces, \sys{} sustains up to 2$\times$ higher request rates than existing T2I workflow serving systems with the same GPU budget; for a fixed request rate, it saves up to 3$\times$ GPU resources while satisfying service level objectives (SLOs).
\end{abstract}

%% file: contents/1_introduction.tex
\section{Introduction}
\def\thefootnote{*}\footnotetext{Equal contribution; $^\#$ Corresponding author}
\label{sec:introduction}

Text-to-image (T2I) workflows built on diffusion models are a cornerstone of
modern image generation~\cite{ootd,
ju2024BrushNet,zhang2023controlnet,gpt4o,dall-e, about_openart}, underpinning
commercial services~\cite{dall-e,midjourney,firefly,katz} that serve millions of
users at more than 10K requests per second (RPS) in
production~\cite{diffusion_production}. In a mainstream public cloud platform,
we observe growing demand for serverless T2I deployments driven by two workload
characteristics. \emph{First}, unlike conventional large language model (LLM)
services exposed through standard APIs~\cite{openai_api}, T2I applications are
highly \emph{customized}. Professional creators compose unique workflows from
diverse diffusion models and adapters (e.g., LoRAs~\cite{hu2022lora}) based on
their application needs; recent Alibaba production
traces~\cite{katz,diffusion_production} reveal 31,133 distinct workflows in
a single 20-day period. \emph{Second}, T2I demand is often \emph{ad hoc} and
\emph{bursty}. Serverless deployment addresses these needs by letting users
upload custom workflows while delegating provisioning, elastic scaling, and
accounting to the platform. Consequently, a serverless T2I platform must
efficiently multiplex customized workflows under dynamic traffic,
all while hiding infrastructure management details like GPU assignment and data
movement from the user.

However, current serverless cloud platforms fall short in T2I serving. As a
common practice, users are required to compose workflows using third-party tools
such as ComfyUI~\cite{comfyUI} or Diffusers~\cite{diffusers}, and then deploy
the entire T2I workflow as a \emph{single, monolithic GPU
function}~\cite{aliyun_create_gpu_function, gcp_serverless_gpu, aws_serverless_gpu}. This practice obscures internal
model executions and data exchanges, leading to three problems that burden
users and limit platform efficiency. 
\emph{First}, while existing works enable multi-GPU parallelism for T2I inference~\cite{katz, li2024DistriFusion, fang2024xdit}, they target limited workflows and lack abstractions for flexible model placement and tensor movement. 
This forces users building customized workflows to manually manage GPU assignment and inter-GPU communication, violating the serverless principle of hiding infrastructure details~\cite{what_serverless_is}.
\emph{Second}, T2I workflows contain large
models of sizes up to tens of GiB, making scale-out bottlenecked by loading
overhead (cold starts~\cite{fork_in_the_road}). Standard techniques that overlap loading with inference
fail when the loading overhead dominates; in a Flux~\cite{flux2024} workflow,
per-model loading is 3.2$\times$--13.5$\times$ slower than the model's single inference pass.
\emph{Third}, GPU scarcity makes multi-tenant T2I serving backlog-prone during
peak hours~\cite{llm_fairness,fc_quota}: on our platform, tens of thousands of
requests can queue, and major users can experience up to 10\% of requests
backlogged. This exposes a \emph{fair scheduling} problem: because user-composed
workflows have heterogeneous resource demands, treating them as uniform requests
allows costlier workflows to consume a disproportionate share of GPU time.

Despite these challenges, serverless T2I inference for customized
workflows remains largely unexplored. Prior systems~\cite{vllm_omni,
sglang_diffusion, fang2024xdit, katz, nirvana, li2024DistriFusion,
lu2026tetriserveefficientditserving} optimize individual T2I workflows through
kernel optimization, parallelization, or in-workflow caching, but do not address the
multi-tenant challenge of serving thousands of distinct workflows under dynamic
traffic. We propose \sys{}, a \emph{serverless-native} system for T2I workflow
inference. \sys{} represents workflows as \emph{model
DAGs} (directed acyclic graphs), making individual model invocations and their
data dependencies the fundamental units for execution, scaling, and scheduling.
\sys{} introduces three system components, each addressing one challenge.

\PHM{Workflow DAG Representation.}
\sys{} exposes a new serverless programming
interface that defines a user-platform contract for customized T2I workflows.
Through this interface, a T2I workflow is converted from a monolithic GPU function into an explicit DAG of \emph{model function} invocations: users implement each model as a \emph{model function} that encapsulates model loading and execution and declares its inputs and outputs; \sys{} uses these declarations to infer the workflow DAG.
This DAG gives \sys{} the data dependency information needed to parallelize independent model functions and materialize intermediate tensors across GPUs, without requiring users to manage GPU placement or inter-GPU communication.
The same abstraction also lets \sys{} control model scaling and fine-grained resource accounting while preserving a familiar serverless programming style, allowing users to focus on the high-level application logic---a key benefit
provided by serverless T2I deployment.

\PHM{GPU-Resident Data Plane.} 
Decomposing a T2I workflow into model functions, while enabling fine-grained
scaling and resource accounting, puts model loading and inter-model data
transfers on the critical path. \sys{} introduces a \emph{unified GPU-resident data
plane} to absorb these costs. Our key insight is that T2I inference is typically
\emph{compute-bound} and leaves substantial GPU memory on the table: in our production
cluster, the P95 GPU memory usage of a diffusion model is 35 GiB, only 36\% of a
modern NVIDIA H20 GPU. 
The system runtime can \emph{harvest} this
unused GPU memory for cached weights and inter-model tensor communication.

\sys{} introduces three mechanisms in the data plane to implement this insight
(\S\ref{sec:unified_data_plane}). \emph{First}, because diffusion-model loading
is often much slower than one inference pass, \sys{} caches the first few layers 
of each model and overlaps loading of the remaining layers with ongoing
inference computation, using a lightweight profiling step to determine the number of layers
needed to hide loading latency. 
\emph{Second}, \sys{} materializes
inter-model dependencies through GPU-resident transfers and callback-based data
fetching, enabling efficient tensor exchange and flexible communication patterns
while remaining transparent to users. 
\emph{Third}, \sys{} manages GPU memory as
a \emph{unified logical address space} shared by active inference, tensor communication, and
cached weights. 
At its core is \emph{model weight virtualization}, which decouples a model's logical weights from their physical placement in GPU memory. 
This enables on-demand allocation of weights across disjoint memory regions. Also, \sys{} treats weights as evictable, spilling them to host memory under pressure.
Together, these mechanisms cut scale-out and communication overhead without exposing the runtime data path to users.

\PHM{Fair Scheduling.} In view of the request backlogs during peak hours and the
need for fair scheduling, \sys{} defines fairness for multi-tenant T2I serving
by the amount of GPU service each tenant receives, rather than by request
counts. For each model function execution, \sys{} measures the GPU time spent on
model loading, computation, and tensor transfer, and charges this cost to the
owning tenant as \emph{vTime}. The scheduler prioritizes tenants with lower
cumulative \emph{vTime}, but strictly enforcing this order can rule out dispatch
choices that would improve serving efficiency, e.g., reduce queuing delay.
\sys{} therefore uses a \emph{two-stage scheduler} that bounds unfairness while
preserving scheduling flexibility. In each scheduling round, the first stage
admits only tenants whose cumulative \emph{vTime} is within an
operator-configured slack of the least-served tenant. The second stage ranks
ready model functions from this eligible set using queuing delay, execution cost, and
remaining workflow work. This design achieves  fairness at \emph{model
function} granularity while allowing workload-aware decisions that improve efficiency.

We prototyped \sys{} and evaluated its decomposition, data plane, and scheduler
under realistic workload variation. Our evaluation spans 20 representative T2I
workflows, including SD3.5~\cite{sd35}, Z-Image~\cite{zimage}, Flux~\cite{flux2,
flux2024}, and various adapters. Using production traces on a multi-GPU testbed,
we compare \sys{} against state-of-the-art serving systems~\cite{vllm_omni,
diffusers}. \sys{} sustains up to 2$\times$ higher request rates at the same GPU
budget, reduces GPU demand by up to 3$\times$ at a fixed rate, and meets up to
7$\times$ tighter SLOs while tolerating 2$\times$ higher burstiness.
Microbenchmarks confirm that the data plane effectively hides loading and
communication overheads, and that the scheduler successfully balances tenant
fairness with throughput.

%% file: contents/2_background.tex
\section{Background and Motivation}
\label{sec:background}

\begin{figure}[t]
  \centering
  \includegraphics[width=0.99\linewidth]{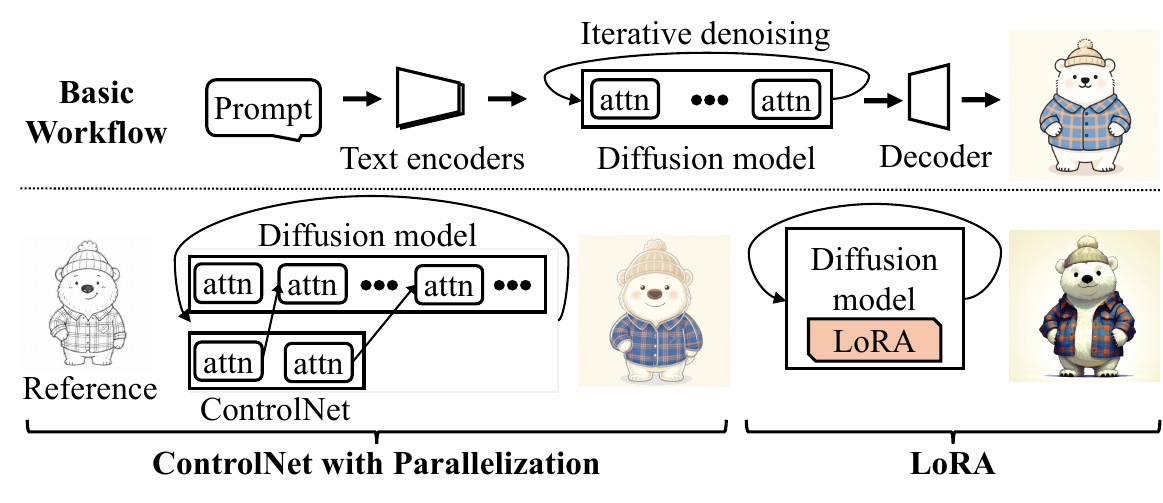}
  \caption{Basic Workflow and Workflows augmented with ControlNet~\cite{zhang2023controlnet} and LoRA~\cite{hu2022lora}.}
  \Description{}
  \label{fig:workflow_example}
  \vspace{-.3in}
\end{figure}

\subsection{A Primer on Text-to-Image Workflow}
\label{sec:primer}

\PHB{Basic Text-to-Image Workflows.} 
As shown in \figref{fig:workflow_example}-top, a \emph{basic} text-to-image generation workflow consists of three types of models: \emph{text encoder}, \emph{base diffusion model}, and \emph{decoder-only variational autoencoder} (VAE). 
The process begins with the text encoder, which encodes a text prompt into a sequence of text embeddings. The system then initializes a latent tensor with random Gaussian noise. Conditioned on the text embeddings, the base diffusion model \emph{iteratively refines} this tensor through a series of \emph{denoising steps}. Finally, the denoised latent representation is passed to the VAE decoder, which reconstructs the output image in pixel space.

\PHB{Adapter-augmented Workflows.}
Production T2I workflows often augment the basic pipeline with \emph{adapter} models to provide fine-grained control over visual attributes such as spatial structure, artistic style, and illumination~\cite{katz,zhang2023controlnet,hu2022lora,zhang2025scaling_iclight,Fooocus,ye2023IP-Adapter}. 
From a systems perspective, two classes of adapters are most relevant, as shown in  \figref{fig:workflow_example}-bottom.

First, \emph{tandem adapters}, such as ControlNet~\cite{zhang2023controlnet}, execute alongside the base diffusion model at each denoising step and inject spatial conditioning signals such as edges or depth maps. 
These adapters complicate serving because their parameter sizes are often comparable to the base model, which increases both model-loading latency and inference latency; moreover, maximizing throughput often requires parallelizing the adapter and base model across GPUs, introducing nontrivial synchronization and data communication~\cite{katz}.

Second, \emph{weight-update adapters}, such as LoRA~\cite{hu2022lora}, modify the base model's weights before inference. 
While they do not invoke an additional model, they introduce loading overhead: as users may rely on many such adapters, they are typically fetched and applied before inference on demand~\cite{katz, diffusion_production}.

\subsection{Serverless T2I Workflows Serving in Production}
\label{sec:current_practice}

T2I workflows are becoming a major serverless workload: in a one-month trace from our platform in March 2026, they account for 28\% of total GPU usage, the largest share among GPU workloads.
We identify two key reasons.

\PHB{Prevalence of Customized T2I Workflows.}
Unlike DNN and LLM services, which are typically exposed through standardized APIs~\cite{openai_api}, T2I applications are often built as customized workflows, especially by professional creators with specialized visual requirements.
These workflows span diverse applications, such as virtual try-on~\cite{ootd_dataset} and image editing~\cite{jiang2026flashps}.
Recent Alibaba production traces~\cite{katz,diffusion_production} show this diversity at scale, with 31,133 distinct workflows recorded over 20 days; we observe similar trends on our platform.

In addition to aesthetic customization, users may incorporate T2I-specific parallelization techniques to accelerate workflow execution, such as ControlNet parallelization~\cite{katz} and sequence parallelism~\cite{katz,li2024DistriFusion,fang2024xdit}.
These techniques exploit the data dependencies within T2I workflows to parallelize adapter execution and base model inference.
For example, in \figref{fig:workflow_example}-bottom, ControlNet produces intermediate results that must be transferred to specific layers of the base model at each denoising step; otherwise, the base model stalls waiting for unavailable inputs.
Exploiting this parallelism therefore requires specialized system support~\cite{katz,distrifusion_controlnet}.

\begin{figure}[t]
  \centering
  \includegraphics[width=0.99\linewidth]{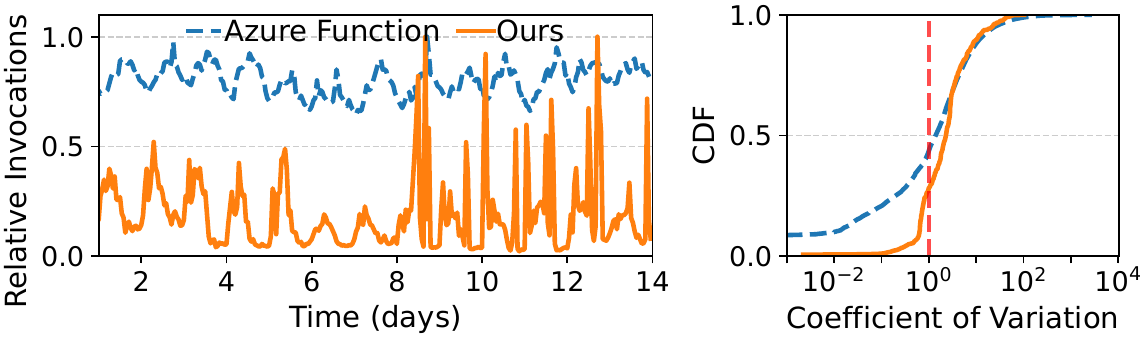}
  \caption{\textbf{Left:} Invocations per hour, normalized to the peak. \textbf{Right:} CDF of CoV values across functions.}
  \Description{}
  \label{fig:normalized_invocations}
  \vspace{-.3in}
\end{figure}

\PHB{Dynamic Workload Traffic.}
\figref{fig:normalized_invocations}-left plots the request volume per hour in our production cluster, normalized by the peak hourly load observed from March 1 to March 14, 2026. 
We include the Azure Functions trace~\cite{azure_function_trace} as a comparison baseline. 
Our T2I workload is visibly more bursty: whereas the Azure trace maintains a relatively stable baseline at roughly 60\% of its peak, the T2I trace fluctuates over a much wider range.

We further quantify workload variability using the coefficient of variation (CoV) of request inter-arrival times (IATs).
A Poisson arrival process has an IAT CoV of 1, while values above 1 indicate burstier arrivals~\cite{azure_function_trace}.
\figref{fig:normalized_invocations}-right shows the distribution of IAT CoV values in our T2I trace and the Azure trace.
Both workloads exhibit substantial request variability: 75\% of functions in the T2I trace have an IAT CoV above 1, compared with 60\% in Azure Functions.

\subsection{Limitations of Current Practices}
\label{sec:limitations}

\enlargethispage{2\baselineskip}

\PHB{Develop and Deploy a T2I Workflow.}
Serverless T2I services are currently provided in the cloud~\cite{aliyun_fc_funart,hf_gpu_space}, where users first build a workflow with existing tools, such as ComfyUI~\cite{comfyUI} or Diffusers~\cite{diffusers}. 
They then provision GPU resources, upload the \emph{entire} workflow as a GPU function~\cite{aliyun_create_gpu_function,hf_gpu_space}, and deploy it in the cloud. 
After that, the platform follows the standard serverless computing paradigm: it manages a shared GPU cluster and automatically scales the GPU functions in response to request traffic. 
Users are billed on a pay-as-you-go basis by accounting GPU usage based on the GPU specification and execution duration~\cite{golgi,aws_lamda_billing,aliyun_fc_billing, hf_gpu_space}.

However, we identify following limitations, each of which undermines an essential property of serverless computing~\cite{what_serverless_is}.

\PHB{L1: High Scaling Overhead.}
Current serverless T2I platforms typically deploy each workflow as a single GPU function~\cite{aliyun_fc_funart, hf_gpu_space}. 
This practice makes the workflow, rather than an individual model component, the unit of scaling. 
When load increases, the platform must therefore replicate the full workflow even if only one component is the actual bottleneck. 
This coarse granularity increases auto-scaling overhead\footnote{Like~\cite{torpor}, we exclude the delay of fetching a remote container image for cold starts, which can take extra seconds to minutes to complete.} in both startup latency and GPU memory footprint. 
For example, scaling a basic Flux1-Schnell workflow on H800 with Diffusers~\cite{diffusers} takes 1.1 seconds even when loading from pinned host memory, adding 110\% overhead relative to model inference under default settings. 
It also consumes 42\% more GPU memory than scaling only the base diffusion model, which is typically the bottleneck. 
We observe similar behavior in vLLM-Omni~\cite{vllm_omni}. 
The root cause is common across existing systems~\cite{vllm_omni, sglang_diffusion, diffusers, fang2024xdit, katz, li2024DistriFusion}: they reuse Diffusers' design~\cite{diffusers}, which follows a ``single-file'' abstraction~\cite{diffusers_philosophy} and packages the \emph{entire} generation workflow, including the base model, adapters, and control logic, as a \emph{monolithic} unit.

\PHB{L2: Exposed Communication Complexity.} 
High performance T2I execution often relies on multi-GPU parallelization, but existing systems force serverless users to handle the resulting communication complexity when composing workflows.
In existing systems~\cite{li2024DistriFusion,katz,fang2024xdit}, the communication logic is tightly coupled to framework internals. 
Adapting these techniques to a user's customized workflow therefore requires substantial systems expertise and engineering~\cite{fang2024xdit,vllm_omni,katz,li2024DistriFusion}. 
In practice, users must reason directly about GPU placement, synchronization, and data movement at runtime~\cite{multi_gpu_comfyui}.
This requirement exposes low-level resource management during workflow development, conflicting with the serverless principle of hiding infrastructure from users~\cite{what_serverless_is}. 
For example, combining ControlNet parallelization with sequence parallelism requires users to understand framework internals, tensor sharding, and low-level distributed communication~\cite{distrifusion_controlnet}, although these details should be hidden behind a serverless abstraction~\cite{pyren}.

\begin{figure}[t]
  \centering
  \includegraphics[width=0.99\linewidth]{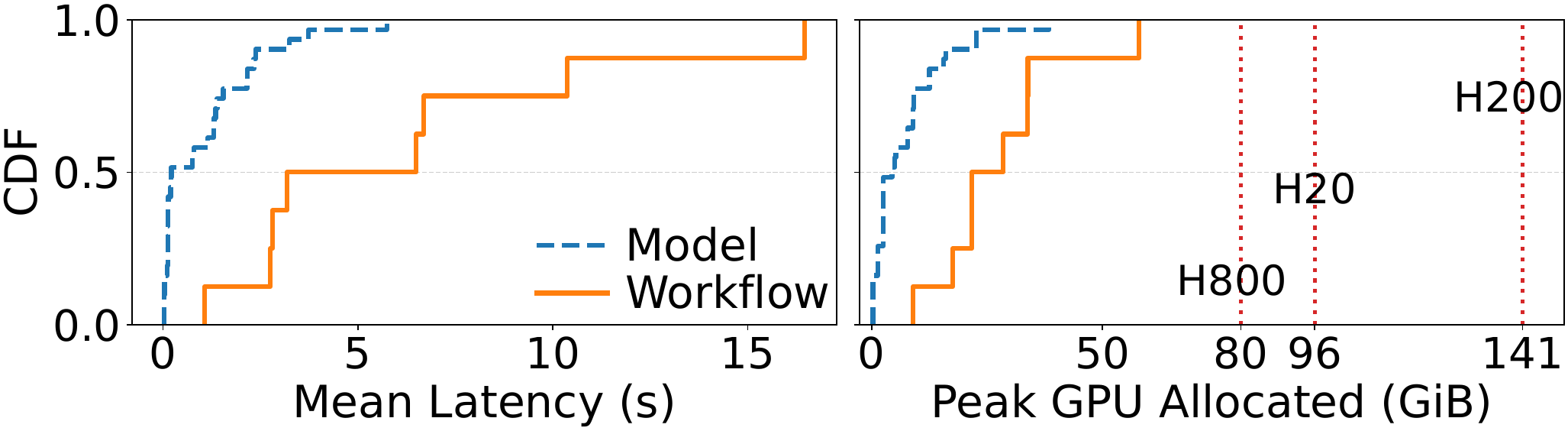}
  \caption{CDF of latency and GPU memory usage of eight workflows\protect\footnotemark[\value{footnote}] on H800 at model level and workflow level.}
  \Description{}
  \label{fig:latency_memory_heterogeneity}
  \vspace{-.3in}
\end{figure}

\PHB{L3: Limited Support for Multi-tenant Serving.} 
In a serverless platform, multiple tenants share a GPU pool to execute their workflows.
However, GPUs are scarce resources, so bursty demand can quickly create request backlogs during peak periods; in our production cluster, tens of thousands of requests can queue, and major users can see up to 10\% of requests backlogged. 
Backlog makes fairness a scheduling requirement: one tenant should not consume a disproportionate share of GPU service and delay others. 
Existing per-tenant quotas, such as request-per-minute (RPM) limits~\cite{fc_quota, openai_rate_limits}, provide isolation but are not work-conserving: they can throttle a tenant even when GPUs are idle.
Worse, quotas account for requests rather than GPU consumption, which mismatches T2I workflows whose costs vary widely: as shown in \figref{fig:latency_memory_heterogeneity}-left, the inference latency of eight popular workflows\footnotemark spans up to 16$\times$. 
While T2I serving systems~\cite{katz,diffusers,fang2024xdit,vllm_omni,sglang_diffusion} can be efficient in single-instance deployments, they largely lack scheduling mechanisms that ensure both fairness and efficiency in a multi-tenant cloud.

\footnotetext{Qwen-Image, Z-Image, Z-Image-Turbo, Flux1-Schnell, Flux1-Dev, SD3, SD3.5, and SDXL with default settings in ~\cite{diffusers}, e.g., resolutions (1024$\times$1024).}

%% file: contents/3_overview.tex
\section{Motivation and System Overview}

\begin{figure}[t]
  \centering
  \includegraphics[width=1.0\linewidth]{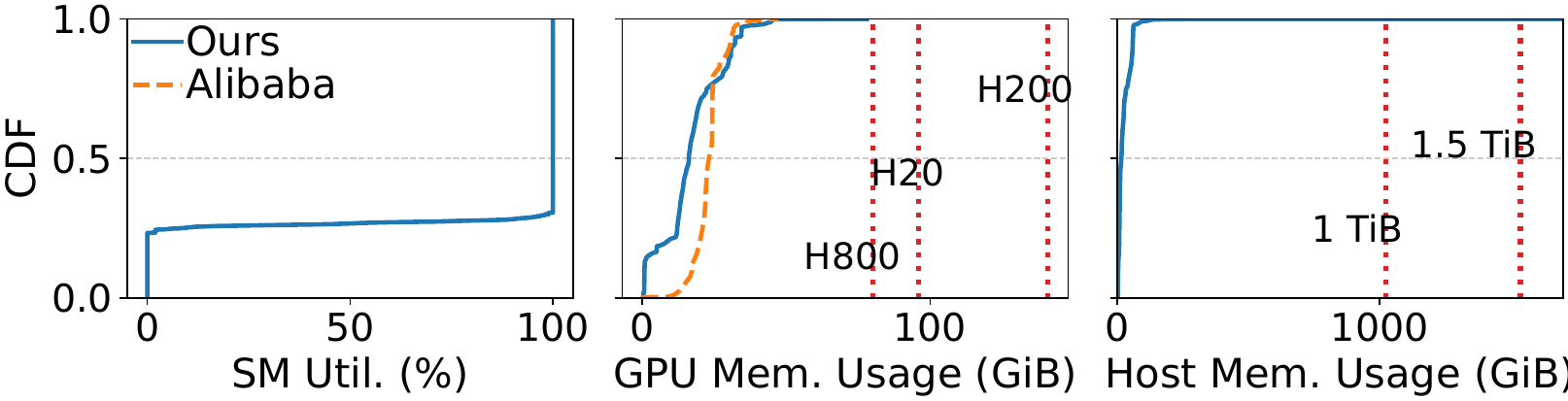}
  \caption{Resource utilization. Recent Alibaba trace~\cite{diffusion_production} profiles GPU memory \emph{but omits SM util. and host memory}. }
  \Description{}
  \label{fig:resource_util}
  \vspace{-.3in}
\end{figure}

\subsection{Key Insight}
\label{sec:key_insight}

\PHB{Compute-bound T2I leaves harvestable GPU memory.}
T2I inference is typically compute-bound: even an inference batch of one can saturate a high-end GPU~\cite{li2024DistriFusion, katz, diffusion_production}. 
Unlike LLM serving, which reserves substantial GPU memory for runtime state such as the KV cache needed to serve a large batch of requests~\cite{vllm}, T2I inference maintains much less runtime state and therefore leaves a substantial fraction of GPU memory unused. 
For example, in Flux workflows, runtime state accounts for only 7.6\% of the total model memory footprint. 
\figref{fig:latency_memory_heterogeneity}-right further shows that both end-to-end workflows and individual diffusion models use only a small fraction of the memory available on a modern GPU.

Meanwhile, we observe the same pattern in our production workloads. 
As shown in~\figref{fig:resource_util}, GPU SM utilization during T2I function execution is high, with the P50 reaching 100\%. 
In contrast, GPU memory usage is much lower: the P50 is only 16~GiB and the P95 is 35~GiB. 
Since modern GPUs provide 80--141~GiB of memory, a large fraction of GPU memory remains idle in production.
Another production trace from Alibaba reports a similar pattern, with P50 and P95 memory usage of 30~GiB and 36~GiB, respectively~\cite{diffusion_production}.
We further observe that host memory is significantly underutilized on these servers, motivating the design described in \S\ref{sec:data_plane_loading}.

\PHB{How does it work in \sys{}?}
\sys{} turns this slack memory into a backend data plane for serverless T2I serving (\S\ref{sec:unified_data_plane}). 
At a high level, the data plane serves two roles. 
First, it caches model weights in otherwise idle GPU memory, reducing model loading overhead for fast scaling.
Second, because \sys{} decomposes a T2I workflow into a DAG of model functions (\S\ref{sec:programming_interface}), the data plane provides GPU-resident buffers for fast data movement along DAG edges. 
This keeps critical-path inter-model communication efficient.
As a benefit of DAG execution, the exposed function boundaries also enable fine-grained resource accounting for fair scheduling (\S\ref{sec:scheduling}).

\subsection{System Overview} 
\sys{} treats a T2I workflow as a model DAG: individual models are exposed as functions, and workflow execution is a sequence of function invocations with data dependencies.

\begin{figure}[t]
  \centering
  \includegraphics[width=0.99\linewidth]{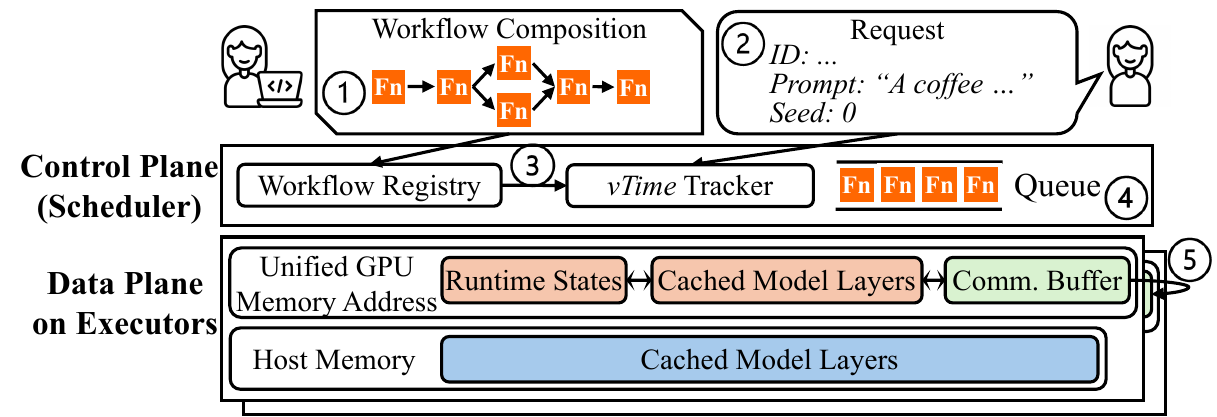}
  \caption{System Overview.}
  \Description{}
  \label{fig:architecture}
  \vspace{-.2in}
\end{figure}

\PHB{System Architecture.}
\figref{fig:architecture} shows the architecture of \sys{}.
At the frontend, users implement model functions, compose them into workflows, and register the workflows with the system (\circledtext{1}). 
\sys{} provides a serverless-style programming interface with T2I-specific abstractions for workflow composition, hiding the complex data movement required by DAG execution from users and addressing \textbf{L2}.
After registration, users invoke a workflow with inputs such as text prompts through OpenAI-compatible APIs~\cite{openai_api} (\circledtext{2}).

At the backend, \sys{} maintains workflow state and uses a fair scheduler to dispatch ready function executions across distributed GPU executors (\textbf{L3}). 
Each executor manages its GPU memory as a unified address space for active model inference runtime, weight caching for fast scaling (\textbf{L1}), and efficient communication between dependent model functions (\circledtext{5}). 
Host memory acts as a secondary storage tier (\S\ref{sec:data_plane_loading}).

\PHB{Life of a Request.}
When a request arrives, the control plane instantiates the corresponding workflow DAG (\circledtext{3}) and tracks the readiness of each model function. 
A \emph{vTime} tracker accounts for each tenant's resource share to support fair scheduling, as described in \S\ref{sec:scheduling}. 
The scheduler selects ready functions, i.e., functions whose inputs are available, and dispatches them to GPU executors (\circledtext{4}). 
After executing a function, the executor reports completion to the control plane and exposes the output through the data plane. 
The control plane then releases downstream functions once their inputs become ready; this process repeats until the workflow output is returned to the user.

%% file: contents/4_design.tex
\section{Programming Interface}
\label{sec:programming_interface}

\PHB{Design Objectives.}
First, \sys{}'s programming interfaces resemble those provided by commercial serverless platforms~\cite{aws_sagemaker_programming,google_cloud_programming,azure_cloud_programming}, which reduces the additional learning curve for users. 
Second, \sys{} abstracts complex data communication behind intuitive APIs, making data communication transparent to users and hiding the underlying infrastructure as well as the complexity of operating it~\cite{what_serverless_is}.

With \sys{}, individual models are implemented as functions and users can use an orchestrator function to assemble them into a workflow in a \emph{declarative} manner. 
\sys{}'s runtime handles execution and communication in the shared resource pool while respecting the workflow's data dependencies.

\PHB{Model Function Development.}
\figref{fig:flux_function_code} illustrates how to implement a Flux1-Dev model in \sys{}. 
Developers implement three methods: $\texttt{setup\_io()}$, $\texttt{load\_model()}$, and $\texttt{execute()}$. 
For $\texttt{setup\_io()}$, \sys{} provides two primitives, $\texttt{add\_input}$ and $\texttt{add\_output}$, to specify the model's input and output interfaces. The other two methods encapsulate model loading and inference, respectively. 
As in prior systems~\cite{sglang_diffusion,vllm_omni,fang2024xdit,katz}, users can reuse model implementation code from Diffusers~\cite{diffusers} when implementing these methods, which also preserves compatibility with inference optimizations such as \texttt{torch.compile()}.

Under the hood, \sys{} organizes the three methods under a \texttt{ModelFn} class, the scaling and management unit at runtime.

\PHB{Assemble Models as a Workflow.}
\sys{} uses an orchestrator function to assemble models as a workflow, similar to the design of Azure Durable function~\cite{azure_durable_orchestrator}.
\sys{} adopts a \emph{declarative} workflow programming model.
The I/O interfaces declared by \texttt{setup\_io} in each \texttt{ModelFn} are sufficient for \sys{}
to parse the data dependencies and apply topological sorting to infer the workflow DAG.
For example, in \figref{fig:workflow_definition}, \texttt{prompt\_embeds} is the output of \texttt{text\_enc}, which serves as the input for \texttt{controlnet} and \texttt{flux}, meaning both of them depend on the output of \texttt{text\_enc} for inference. 
At runtime, \sys{} materializes the data communication while hiding the implementation details from users, as elaborated in \S\ref{sec:data_plane_communication}.

\PHB{Function-as-a-Service Paradigm.}
In \sys{}, \texttt{ModelFn} corresponds to the notion of a function in the Function-as-a-Service paradigm. 
It abstracts model components in a T2I workflow as small, single-purpose functions that are exposed to the serverless platform, which scales them on demand. 
\texttt{ModelFn} is also stateless and with random seeds fixed, the same method inputs always produce the same outputs, allowing safe retries for fault tolerance.

\begin{figure}[tb]
\begin{lstlisting}[style=mypython, escapechar=|]
## Developers start here ##
def setup_io(self) -> None:
  # define inputs
  add_input("latents")
  add_input("prompt_embeds")
  add_input("control_outputs", callback=True)
  # define outputs
  add_output("noise_pred")
        
def load_model(model_path):
  transformer = FluxTransformer2DModel.from_pretrained(
    model_path, torch_dtype=torch.bfloat16,
  )
  return {"transformer": transformer}

@torch.no_grad()
def execute(model_components, **kwargs):
  transformer = model_components["transformer"]
  noise_pred = transformer.forward(**kwargs)
  return {"noise_pred": noise_pred}

## Invisible from developers ##
class ModelFn:
  setup_io = setup_io
  initialize = load_model
  execute = execute
\end{lstlisting}
\vspace{-.3in}
\caption{A simplified implementation of Flux function.}
\Description{}
\label{fig:flux_function_code}
\vspace{-.2in}
\end{figure}

\begin{figure}[tb]
\begin{lstlisting}[style=mypython, escapechar=|]
def compose_workflow():
  # create model function instances
  text_enc = ModelFn(model_path=model_path)
  flux = ModelFn(model_path=model_path)
  controlnet = ModelFn(model_path=controlnet_path)
  vae = ModelFn(model_path=model_path)
  ...
  # connect function instances
  latents = random_latents_generator(seed)
  prompt_embeds = text_enc(prompt)
  for i in range(num_steps): # iterative denoising
    control_outputs = controlnet(latents, prompt_embeds)
    noise_pred = flux(latents, prompt_embeds, control_outputs)
    latents = denoise(noise_pred, latents) 
  output_img = vae(latents, mode="decode")
  ...
\end{lstlisting}
\vspace{-.3in}
\caption{A simplified Flux workflow composition.}
\Description{}
\label{fig:workflow_definition}
\vspace{-.2in}
\end{figure}

\section{Data Plane}
\label{sec:unified_data_plane}
\sys{}'s data plane builds on the insight in \S\ref{sec:key_insight}: harvested GPU memory can reduce two costs on the critical path, model loading and data movement across functions.
We describe how to reduce model loading overhead (\S\ref{sec:data_plane_loading}) and provide efficient data communication (\S\ref{sec:data_plane_communication}), then explain how a unified logical address space manages GPU memory (\S\ref{sec:data_plane_unified}).

\subsection{Model Loading}
\label{sec:data_plane_loading}

\PHB{Challenge: Model loading remains on the critical path.}
Decomposing a T2I workflow into model functions gives the platform control over each model, but it does not remove model loading during scale out.
As workflows have large models along the critical path, loading remains a  bottleneck.

Prior systems reduce loading overhead by overlapping model loading with execution in two ways. 
One approach transfers later layers of a model while its earlier layers are executing~\cite{pipeswitch,toppings,torpor,jiang2026flashps}. 
Another approach loads later stage models in a workflow while earlier stage models are executing~\cite{orion}. 
Both approaches require enough execution time to hide loading, which often does not hold for T2I workflows.
For example, in a basic Flux1-Dev workflow, loading \texttt{text\_encoder}, \texttt{text\_encoder\_2}, \texttt{transformer}, and the \texttt{VAE} decoder from pinned host memory to an NVIDIA H800 GPU takes 5$\times$, 13.5$\times$, 3.2$\times$, and 6.3$\times$ their respective inference latencies.
Even with ideal overlap, layer by layer loading reduces Flux1-Dev end-to-end latency by only 13\%.
Prewarming later stage models also leaves a large bottleneck: even with weights cached in host memory, model loading still accounts for 44\% of end-to-end latency.
Prewarming further interferes with multi-tenant scheduling by injecting bursts of loading traffic that may delay other tenants.

\PHB{Caching the first layers.}
\sys{} reduces model loading overhead with a multi tier cache spanning GPU memory, host memory, and external storage.
GPU memory is the fastest tier, caching model weights for immediate execution.
Host memory is the second tier, holding weights that can be loaded into GPU memory on demand.
External storage is the last tier with nearly unlimited capacity but much lower bandwidth.
Consistent with prior observations~\cite{diffusion_production,katz}, workflow popularity is skewed. 
In our trace replay, caching the 15 most popular workflows in underutilized host memory serves 99\% of model loads from host memory, as detailed in Appendix~\ref{sec:host_memory_loading}. 
Accordingly, we focus on host to GPU loading.

\begin{figure}[t]
  \centering
  \includegraphics[width=0.99\linewidth]{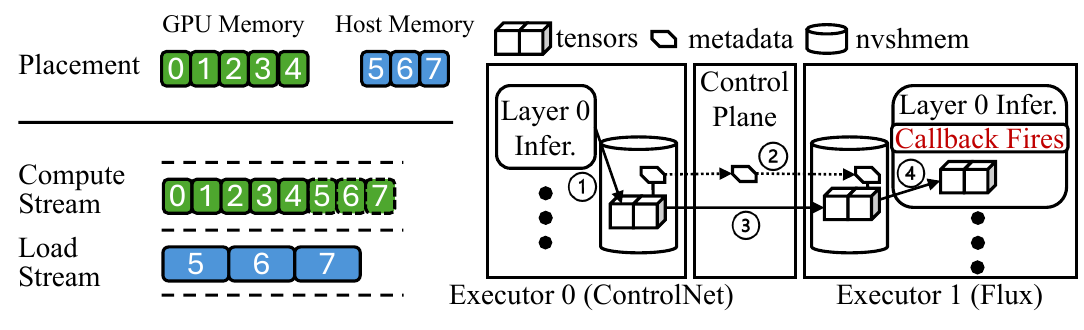}
  \caption{\textbf{Left:} A simplified illustration of \sys{}'s loading mechanism. \textbf{Right:} An example of data fetching.}
  \Description{}
  \label{fig:loading_communication}
  \vspace{-.2in}
\end{figure}

The key idea is to split each model across the two tiers:  cache its first layers in GPU memory and keep the remaining layers in host memory, as shown in \figref{fig:loading_communication}-left.
Consider a model with $L$ uniform layers.
\sys{} preloads the first $L_{e}$ layers into GPU memory and keeps the remaining $L - L_{e}$ layers in host memory.
When a request arrives, inference starts immediately on the cached first layers, while loading   the remaining layers from host memory asynchronously.
Let $T_{load}$ denote the loading latency of one layer and $T_{comp}$ denote its inference latency.
Because $T_{load} \gg T_{comp}$, the end-to-end latency $T_{e2e}$ as a function of $L_{e}$ is
\begin{align}
\label{eq:e2e_latency}
    T_{e2e}(L_{e}) = \max (LT_{comp}, (L-L_{e})T_{load} + T_{comp}).
\end{align}

Eq.~\ref{eq:e2e_latency} gives the minimum prefix length $L_{e}^{\prime}$ required to fully hide model loading:
\begin{align}
\label{eq:l_head}
L_{e}^{\prime}=\left\lceil \frac{L\!\left(T_{\mathrm{load}}-T_{\mathrm{comp}}\right)+T_{\mathrm{comp}}}{T_{\mathrm{load}}} \right\rceil .
\end{align}

Computing $L_{e}^{\prime}$ requires only two profiled quantities: the per-layer loading latency $T_{load}$ and the per-layer inference latency $T_{comp}$.
For diffusion models and their ControlNets, profiling is lightweight because they are typically composed of uniform transformer blocks with similar costs.
Eq.~\ref{eq:e2e_latency} and Eq.~\ref{eq:l_head} can be extended to models with non-uniform layers.

\subsection{Communication}
\label{sec:data_plane_communication}

\PHB{Challenge: Model DAG execution requires efficient and flexible tensor communication.}
Once a workflow is decomposed into functions, intermediate tensors must be communicated across functions. 
This poses two challenges. 
First, each workflow execution transfers a large volume of data---on the order of GiBs~\cite{katz}---and 99\% of the transferred objects are tensors. 
Second, specialized parallelization techniques introduce complex communication patterns in which both correctness and performance depend on consuming data at the right time. 
For example, if computation blocks while waiting for data to be produced, the system may lose much of the performance benefit from parallel execution (\S\ref{sec:background}).

Existing serverless data planes are a poor fit for T2I workflow execution. 
Data planes that rely on host memory~\cite{pheromone} perform poorly for CUDA tensors: the communication latency for an SD3 \texttt{ModelFn} is 28$\times$ its execution latency due to PCIe transfers, serialization, and socket overhead. 
Recent work builds data planes atop high-speed interconnects such as NVLink and RDMA~\cite{grouter,faasscale}, but exposing these links through collective-style APIs introduces multiple synchronization points to preserve correctness. 
In our measurements, these synchronization points add up to 20\% inference latency when running Flux1-Schnell~\cite{flux2024}, a computation-intensive model, on two NVIDIA H800 GPUs, in both the basic workflow and the ControlNet-augmented workflow.

\PHB{Efficient and flexible data fetching.}
\sys{} combines NVSHMEM~\cite{nvshmem} with callback-based fetching to provide efficient and flexible GPU communication. 
NVSHMEM is a natural substrate for \sys{} because it provides one-sided GPU communication over NVLink and RDMA, supports GPU-initiated transfers, and exposes a symmetric heap abstraction that avoids explicit remote-address management. 
In \sys{}, each executor reserves a fixed-size NVSHMEM arena on its GPU using \texttt{nvshmem\_malloc}\cite{nvshmem_api} and manages the resulting symmetric heap with a buddy allocator\cite{buddy_alloc,buddy_alloc_2}, returning device pointers usable by both local CUDA kernels and remote NVSHMEM operations.  
On top of this substrate, \sys{} allows users to mark a \texttt{ModelFn} input as a \emph{callback} when the input is needed only partway through execution. 
For example, line 6 in \figref{fig:flux_function_code} declares \texttt{control\_outputs} with \texttt{callback=True} to support ControlNet parallelization (\S\ref{sec:primer}). 
The runtime wraps this input and fires the callback when downstream models consume it, overlapping communication with earlier computation. 
Compared with manually managed peer-memory access~\cite{nvidia_peer_device_memory_access}, this design better matches T2I workflows, where intermediate tensors are produced and consumed at fine-grained, model-dependent points during inference.

At runtime, a producer writes its output tensor into local NVSHMEM and publishes the tensor metadata, such as its pointer and shape, to downstream consumers.
When a consumer first needs the tensor, \sys{} allocates a destination block from the consumer's local NVSHMEM arena and pulls the tensor bytes directly from the producer's remote address using a one-sided NVSHMEM operation.
The data path stays in GPU memory, avoiding PCIe round trips and socket communication.
The callback abstraction controls when a transfer occurs: the fetch is issued only when execution reaches the program point where the tensor is needed.

\figref{fig:loading_communication}-right presents an example of data fetching in ControlNet parallelization, where ControlNet execution is interleaved with the base Flux model, as shown in \figref{fig:workflow_example}-bottom.
At runtime, the output of ControlNet
layer~0 is consumed partway through Flux layer~0.
Rather than blocking until
this output is available, Executor~1 begins executing Flux layer~0 and registers
a fetch callback.
When Flux reaches this point,
the callback fires. By then, Executor~0 has produced the ControlNet layer~0
output and placed it in its NVSHMEM communication buffer (\circledtext{1}); the
corresponding tensor metadata is forwarded to Executor~1 (\circledtext{2}),
which uses it to issue a one-sided NVSHMEM fetch into its local tensor store
(\circledtext{3}) before Flux consumes the tensor (\circledtext{4}).
This callback-based fetch path allows Flux to overlap its computation with
ControlNet execution. 
Without it, Flux would have to wait until the relevant
ControlNet output was materialized, eliminating the parallelism between the two
models.

Note that tensor metadata is tiny, on the order of KiB, and executors piggyback it on completion notifications, allowing the control plane to track tensors with little overhead.

\begin{figure*}[t]
  \centering
  \includegraphics[width=1.0\linewidth]{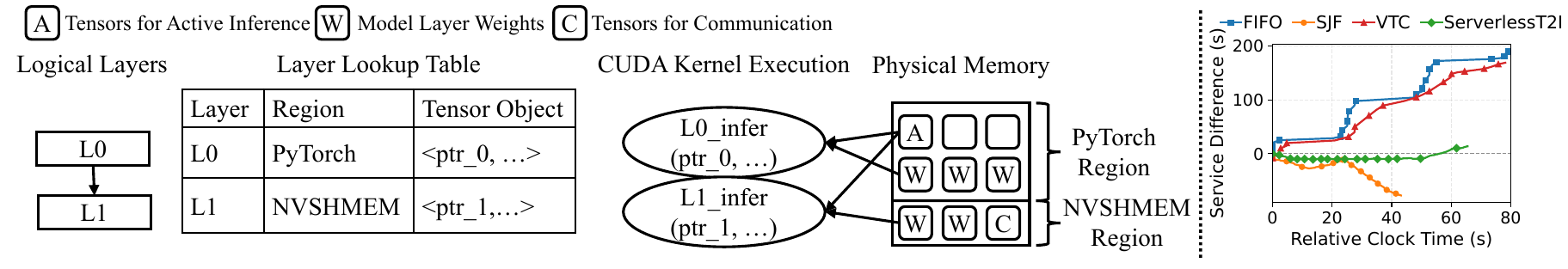}
  \caption{\textbf{Left:} An example of model weight virtualization. \textbf{Right:} Difference of service received for two backlogged tenants.}
  \Description{}
  \label{fig:virtual_memory_fairness}
\end{figure*}

\subsection{Unified GPU Memory Management}
\label{sec:data_plane_unified}
The shared use of GPU memory requires each executor to coordinate three memory consumers within its local GPU: active inference state, communication buffers  (\S\ref{sec:data_plane_communication}), and model weights (\S\ref{sec:data_plane_loading}). 
Active inference state, such as inputs and intermediate activations, is framework-specific and allocated through the deep learning framework runtime, i.e., PyTorch in \sys{}. 
In contrast, communication buffers are allocated and managed by NVSHMEM. 
As a result, local GPU memory is divided into two disjoint allocation domains, which we refer to as the PyTorch region and the NVSHMEM region.

\sys{} introduces \emph{model weight virtualization}, a software indirection layer that decouples a model's logical weights from their physical placement in GPU memory. 
Similar to how virtual memory in operating systems decouples a process's logical address space from physical memory frames, \sys{} decouples a model's logical layers from the physical GPU memory that stores their weights. 
Model weights can transparently reside in either the PyTorch region or the NVSHMEM region, while model execution accesses them through the same logical interface. 
This indirection bridges the two otherwise isolated regions into a unified pool for cached weights, exploiting the underutilized GPU memory (\S\ref{sec:key_insight}).

\PHB{How does it work?}
We walk through an example in \figref{fig:virtual_memory_fairness}-left, to show how \sys{} executes model inference and manages GPU memory. 
From the perspective of model execution, the model still consists of a conventional sequence of \emph{logical layers}, as seen by the PyTorch \texttt{forward()} pass. 
Before executing each layer, \sys{} consults a \emph{Layer Lookup Table} to translate the logical layer into the corresponding tensor objects that represent its weights. 
Each tensor object contains the necessary metadata, such as shape and data type, together with a data pointer to the underlying physical GPU memory. 
This physical memory may be allocated from either the PyTorch region or the NVSHMEM region. 
The CUDA kernels invoked by the framework then execute normally, using the data pointers of both active inference state and model weights to access the appropriate physical memory.

Weights allocated by PyTorch are ordinary tensor objects and can be used directly, whereas weights stored in the NVSHMEM region require pointer rebinding. 
For each NVSHMEM allocation that holds a layer's weights, \sys{} wraps the device pointer as a CUDA tensor using PyTorch C++'s \texttt{torch::from\_blob}, creating a tensor view over the NVSHMEM-backed memory. 
\sys{} then rebinds the model's parameter entries to these tensors and records them in the Tensor Object entries of the \emph{Layer Lookup Table}. 
Consequently, forward() calls access NVSHMEM-resident weights as ordinary PyTorch tensors, and CUDA kernels treat them identically to tensors allocated by PyTorch's CUDA allocator.

\sys{} tracks model weights at layer granularity and treats cached weights as elastic state. 
When GPU memory pressure arises, for example because a new model must be loaded from host memory, \sys{} evicts cached weights layer by layer until the demand is satisfied. 
An executor orders its GPU-resident models by recency and reclaims weights from the least recently used (LRU) models first, evicting layers in reverse layer order to preserve first layers that enable overlapping loading with inference. 
This eviction policy trades a modest latency increase for reclaimed memory while striving to preserve cached weights for fast scale-out.

\PHB{Can NVSHMEM region be elastic?}
The NVSHMEM region size should be specified at initialization~\cite{nvshmem_env}, and resizing it at runtime is impractical because NVSHMEM maintains a symmetric heap with identical size and layout across all GPUs, making any adjustment require cluster-wide coordination~\cite{nvshmem_memory_management}. 
This creates a fundamental tension: an undersized NVSHMEM region risks deadlock when in-flight operations cannot allocate space for data communication, stalling the entire system; an oversized region, however, squeezes the space available for cached weights without \emph{model weight virtualization}, which we quantitatively analyze in \S\ref{sec:microbenchmark}.

\PHB{Backend-as-a-Service Paradigm.}
\sys{}'s data plane follows the Backend-as-a-Service paradigm in serverless computing~\cite{what_serverless_is}.
It exposes model caching and data communication as managed backend services, allowing users to benefit from them without building or operating these components themselves.
In T2I workflow execution, all intermediate data is \emph{immutable}: intermediate tensors produced are consumed once and never
updated~\cite{katz,diffusers}, which obviates consistency protocols and
simplifies fault tolerance.
The data plane reclaims tensors when no downstream \texttt{ModelFn} requires them.
We use \texttt{expandable\_segment}~\cite{expandable_segment} in the PyTorch region and our buddy allocator in the NVSHMEM region to mitigate memory fragmentation.
If an executor fails, \sys{} reconstructs lost data
by re-executing the affected \texttt{ModelFn}s and loading model weights, following a similar approach to prior cluster computing frameworks~\cite{ray, spark, pheromone}.

\section{Ensure Fairness in Multi-tenant Serving}
\label{sec:scheduling}

Resource fairness is critical in multi-tenant serving systems because GPUs are scarce cloud resources, and tenant requests can remain backlogged during peak demand~\cite{llm_fairness}.
Yet existing T2I serving systems provide limited support for resource-aware scheduling~\cite{vllm_omni, sglang_diffusion, fang2024xdit, katz, nirvana, li2024DistriFusion, lu2026tetriserveefficientditserving}.
This gap is especially problematic for T2I workloads: because workflows differ substantially in GPU demand (\figref{fig:latency_memory_heterogeneity}-left), two backlogged tenants that submit requests at the same rate may still receive very different amounts of service.

We illustrate this effect by augmenting Diffusers~\cite{diffusers}, a representative T2I serving system, with three scheduling policies: FIFO, SJF, and an adapted version of VTC~\cite{llm_fairness}, a fairness-oriented scheduler originally designed for LLM serving.
We run the experiment on a four-H800 testbed with two tenants, each issuing requests at 2 RPS.
One tenant invokes a basic SD3-medium~\cite{sd3} workflow, while the other invokes a basic Flux1-Dev~\cite{flux2024} workflow.
We define the service received by a tenant as its cumulative GPU time.

\figref{fig:virtual_memory_fairness}-right reports the service difference between the two tenants over the interval in which both remain backlogged.
None of the policies provides satisfactory fairness.
FIFO accounts for requests rather than service: despite \emph{identical} request rates, the Flux1-Dev tenant receives substantially more GPU time because each Flux1-Dev request is much more expensive, with 10$\times$ the inference latency of SD3-medium.
SJF exhibits the opposite bias, giving more service to the SD3-medium tenant because it consistently favors shorter requests.
Adapting VTC reduces the imbalance by counting T2I workflow operations, analogous to its use of decoding steps in LLM serving~\cite{llm_fairness}, but operation counts remain an inaccurate proxy for GPU service because T2I models differ widely in per-operation cost (\figref{fig:latency_memory_heterogeneity}-left).

\PHB{Fine-grained Fairness Accounting with \emph{vTime}.}
\sys{} accounts for service at the granularity of each \texttt{ModelFn} execution. 
Whenever a \texttt{ModelFn} runs on a GPU, \sys{} charges its owner a \emph{vTime} equal to the wall-clock GPU time consumed, including computation, model loading, and tensor transfer. 
At dispatch time, the scheduler charges an estimated \emph{vTime} derived from historical measurements so that scheduling decisions can proceed immediately; once execution completes, the estimate is replaced with the measured GPU time, keeping accounting faithful to realized usage. 
If a denoising step launches parallel \texttt{ModelFn}s (e.g., a base model and a ControlNet), the tenant is charged for both.

\PHB{Fairness Scheduling with \emph{vTime}.}
\sys{} tracks the cumulative \emph{vTime} of each tenant and prioritizes those that have received less service. 
For a tenant that newly enters the system or returns after inactivity, \sys{} performs a \emph{vTime lift}, initializing its vTime to the minimum among active tenants. 
Without this lift, a returning tenant's stale, artificially low \emph{vTime} would grant it repeated priority until it catches up, converting past absence into a scheduling advantage.

To balance fairness and serving efficiency, 
\sys{} allows a configurable slack so that the scheduler retains scheduling flexibility. 
This is realized as a two-layer scheduler.

\textbf{\textit{Layer~1: Fairness filter.}}
The scheduler maintains cumulative \emph{vTime} $v_u$ for each tenant $u$.
At dispatch time, it forms a \emph{candidate set} $C$ of tenants:
a tenant $u$ is included if $v_u - \min_{u'} v_{u'} \leq \Delta$, where $\Delta$ is an operator-configured slack.
Only \texttt{ModelFn}s belonging to tenants in $C$ proceed to Layer~2.
A smaller $\Delta$ enforces stricter fairness, while a larger $\Delta$ gives Layer~2 a wider pool of candidates to optimize throughput.

\textbf{\textit{Layer~2: Priority scoring.}}
Strict fairness alone can hurt serving efficiency.
For example, the scheduler may interleave the execution of two workflows to ensure fair resource allocation, increasing the latency of both.
To improve serving efficiency,
among the eligible \texttt{ModelFn}s, Layer~2 ranks each \texttt{ModelFn}~$n$ by a normalized priority
score
$\textit{score}(n) = \hat{w}(n) - \hat{r}(n) - \hat{e}(n)$,
where $\hat{w}(n)$, $\hat{r}(n)$, and $\hat{e}(n)$
denote its waiting time, the remaining critical-path time of its workflow, and its execution latency, respectively.
The scheduler dispatches the highest-scoring \texttt{ModelFn}.
The terms $-\hat{r}(n)$ and $-\hat{e}(n)$ favor \texttt{ModelFn}s with shorter remaining work or lower execution latency.
The term $\hat{w}(n)$ prevents starvation.
The selected \texttt{ModelFn} is dispatched to an executor that can run it more efficiently, e.g., one with its model already cached to reduce loading overhead or its inputs available in the local GPU memory to reduce communication overhead.

Algorithm~\ref{alg:fair_scheduler} summarizes the scheduling process,
including \emph{vTime} lift, candidate-set construction, priority scoring, 
and \emph{vTime} correction after execution.

\begin{algorithm}[t]
\footnotesize
\DontPrintSemicolon
\caption{Scheduling Algorithm}
\label{alg:fair_scheduler}
\KwIn{Per-tenant cumulative \emph{vTime} $v_u$; fairness slack $\Delta$}

\SetKwProg{Fn}{Function}{:}{}
\SetKwFunction{Schedule}{Schedule}
\SetKwFunction{OnComplete}{OnComplete}

\Fn{\Schedule{}\tcp*[f]{invoked when an executor is idle}}{

\tcp{vTime lift for new/returning tenants}
\ForEach{tenant $u$ with ready \texttt{ModelFn}s}{
  \lIf{$u$ is new or returning}{$v_u \gets \min_{u' \in \mathrm{Active}} v_{u'}$}
}

\tcp{Layer 1: fairness filter}
$R \gets$ tenants with ready \texttt{ModelFn}s\;
$v_{\min} \gets \min_{u \in R}\, v_u$\;
$C \gets \{u \in R \mid v_u - v_{\min} \leq \Delta\}$\;

\tcp{Layer 2: priority scoring}
$N \gets$ ready \texttt{ModelFn}s owned by tenants in $C$\;
$n^\star \gets \arg\max_{n \in N}\; \hat{w}(n) - \hat{r}(n) - \hat{e}(n)$\;

\tcp{Dispatch and charge estimated vTime}
$\hat{T}_{n^\star} \gets$ estimated execution time of $n^\star$\;
$v_{\mathrm{owner}(n^\star)} \mathrel{{+}{=}} \hat{T}_{n^\star}$\;
Dispatch $n^\star$ to best-fit executor\;
}
\Fn{\OnComplete{$n,\, T$}\tcp*[f]{$T$: measured GPU time of $n$}}{
  $v_{\mathrm{owner}(n)} \mathrel{{+}{=}} T - \hat{T}_{n}$ \tcp*{correct vTime}
}
\end{algorithm}

\textbf{\textit{Fairness bound.}}
Since \sys{} serves DAG-structured workflows, a tenant may have outstanding requests but no schedulable work when all pending \texttt{ModelFn}s are blocked on predecessors. 
We thus call a tenant \emph{eligible-backlogged} over $[t_1,t_2)$ if it has at least one ready \texttt{ModelFn} at every point in the interval.  
This refines the backlogged condition in VTC for LLM serving~\cite{llm_fairness}, which only requires a queued request.
This distinction is necessary for DAGs, where dependencies can leave outstanding work with no ready computation.

Let $L_\text{max}$ denote the largest \emph{vTime} charge of any single dispatch, and define $U = \Delta + L_\text{max}$.
Consider any two tenants $f$ and $g$ that remain eligible-backlogged during $[t_1, t_2)$.
For tenant $u$, let $W_u(t_1,t_2) = v_u(t_2) - v_u(t_1)$ denote the service it receives during this interval.
Then \sys{} guarantees
\begin{equation*}
\begin{aligned}
  |W_f(t_1,t_2) - W_g(t_1,t_2)| 
    &\leq |v_f(t_2) - v_g(t_2)| + |v_f(t_1) - v_g(t_1)| \\
    &\leq 2(\Delta + L_{\text{max}}) = 2U.
\end{aligned}
\end{equation*}

%% file: contents/5_implementation.tex
\section{Implementation}
We have implemented \sys{} with a FastAPI~\cite{fastapi} frontend, which exposes a programming interface for users to compose and
register T2I workflows (\S\ref{sec:programming_interface}). 
Users 
invoke their workflows with image generation parameters, such as
prompts and reference images, similar to the OpenAI API~\cite{openai_api}. 
\sys{}'s backend runtime
consists of a control plane and distributed executors (\figref{fig:architecture}),
totaling 4,000 lines of Python code. 
The data plane is implemented in 3,000 lines of Python and C++/CUDA code, built on PyTorch and NVSHMEM~\cite{nvshmem}. 
Aside from CUDA tensors,
communication between the scheduler and distributed executors is facilitated
via ZeroMQ~\cite{zmq}.

%% file: contents/6_evaluation.tex
\section{Evaluation}
We evaluate \sys{} with the following highlights:

\begin{itemize}[topsep=3pt, leftmargin=*, noitemsep, nolistsep, parsep=0pt, partopsep=0pt]
  \item \sys{} outperforms state-of-the-art baselines in controlled end-to-end evaluation, sustaining
  up to 2$\times$ higher request rates, satisfying 7$\times$ more stringent SLOs,
  reducing GPU requirements by up to 3$\times$, or tolerating 2$\times$
  higher burst traffic, all while maintaining 90\% SLO attainment
  (\S\ref{sec:e2e_performance}).

  \item Our microbenchmarks isolate the benefits of \sys{}'s designs: it reduces model loading overhead, enables efficient parallelization with minimal effort, and explores a tradeoff between fairness and efficiency.
  (\S\ref{sec:microbenchmark} \& \S\ref{sec:fair_scheduler}).

  \item \sys{}'s DAG execution adds negligible overhead (\S\ref{sec:system_overhead}).
\end{itemize}

\subsection{Setup}
\label{sec:eval_setup}

\PHM{Testbed and Workloads.} 
By default, we 
use a testbed of 32 NVIDIA H800 GPUs and a scaled real-world T2I production trace collected from our production cluster (\figref{fig:normalized_invocations}-left). 
To evaluate under diverse conditions, we vary request rates, SLO targets, traffic burstiness, and testbed sizes, covering a broad range of traffic patterns and performance requirements.

\PHM{Metrics.}
Our primary metric is \textit{SLO attainment}: the fraction of
requests completed within their specified latency
deadline. 
We set the default deadline to 3$\times$ the solo inference latency of each workflow (SLO Scale $= $3).
Unlike prior works~\cite{katz, li2024DistriFusion, nirvana, fang2024xdit},
\sys{} does not alter the computation of T2I inference and we have validated the identity of generated images.

\PHM{Baselines.}
We primarily compare \sys{} with \vllm{}  and \diffusers{} , which are representative state-of-the-art T2I serving systems~\cite{diffusers, diffusers_server, vllm_omni}.
Following current practices, we deploy each workflow as a \emph{monolithic} GPU function (\S\ref{sec:current_practice}).
Since these systems were originally designed as standalone inference frameworks, we adapt them to the serverless setting and evaluate three 
deployment variants:

\begin{itemize}[topsep=3pt, leftmargin=*, noitemsep, nolistsep, parsep=2pt, partopsep=0pt]

  \item \noCache ({\underline{No} caching}) executes each request to a workflow in a GPU function without any caching on GPU.

  \item \wCache (\underline{W}orkflow cache) utilizes slack GPU memory to cache entire workflows. 
  A GPU function is terminated when the cached workflow is evicted.

  \item \mCache (\underline{M}odel cache) utilizes slack GPU memory to cache individual models within workflows. Because caching is performed at a finer model granularity, this variant can accommodate more workflows in GPU memory.
\end{itemize}

\noindent Note that \wCache and \mCache are augmented versions of \noCache that utilize slack GPU memory and use the same LRU caching policy as \sys{} for a fair comparison.

\PHB{Workflows and Settings.} We compose T2I workflows 
using six popular base models: SD3.5-Large~\cite{sd35}, Z-Image~\cite{zimage}, Z-Image-Turbo~\cite{zimage},
Flux1-Dev~\cite{flux2024}, Flux1-Schnell~\cite{flux2024}, and Flux2-Klein~\cite{flux2}. They exhibit diverse
computational characteristics, with parameter counts spanning 6B to 12B and
denoising steps ranging from 4 to 50. 

We set up two settings, as detailed in \tabref{tab:eval_settings}, randomly assigning workflows to the top-tier tenant traffic from our production trace.
In \textbf{S1}, baselines are based on \vllm{}, as it provides little support for adapters~\cite{vllm_omni} and does not support Flux1-Schnell.
In \textbf{S2}, baselines use \diffusers{}.

\begin{table}[t]
  \centering
  \footnotesize
  \def\arraystretch{0.85} 
  \caption{\textbf{Evaluation settings:}
  \textbf{S1} includes basic workflows, where each workflow consists of text encoders, a diffusion model, and a decoder.
  \textbf{S2} further includes adapter-augmented workflows, which extend the basic workflows with different ControlNet and LoRA adapters.}
  \label{tab:eval_settings}
  \begin{tabular}{clc}
    \toprule
    \textbf{Setting} & \textbf{Diffusion Models} & \textbf{No. Workflows} \\
    \cmidrule(l){1-3}
    S1 & \begin{tabular}[c]{@{}l@{}}
           SD3.5-Large, Z-Image, Z-Image-Turbo \\
           Flux1-Dev, Flux2-Klein
         \end{tabular} & 5 \\
    \cmidrule(l){1-3}
    S2 & \begin{tabular}[c]{@{}l@{}}
           SD3.5-Large, Z-Image, Z-Image-Turbo \\
           Flux1-Dev, Flux1-Schnell, Flux2-Klein
         \end{tabular} & 20 \\
    \bottomrule
  \end{tabular}
  \vspace{-.2in}
\end{table}

\subsection{End-to-end Evaluation}
\label{sec:e2e_performance}

\begin{figure*}[t]
  \centering
  \includegraphics[width=1.0\linewidth]{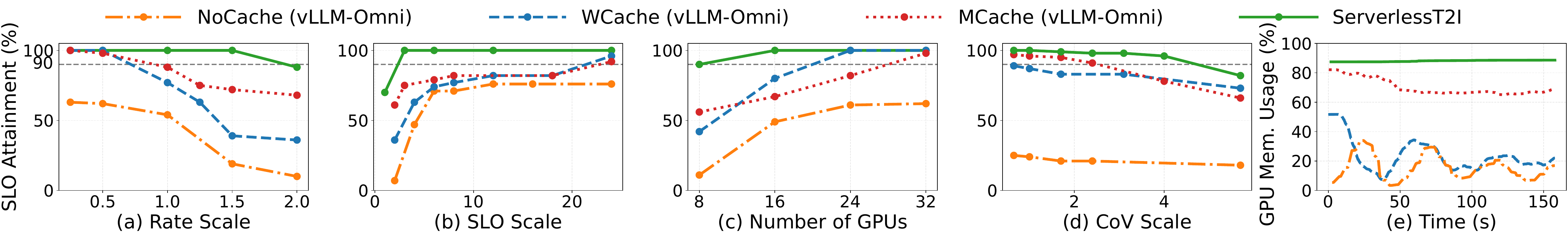}
  \caption{End-to-end evaluation of Setting 1 (\textbf{S1}). All baselines are implemented on top of vLLM-Omni~\cite{vllm_omni}.}
  \Description{}
  \label{fig:s1_end2end}
  \vspace{-.2in}
\end{figure*}

\begin{figure*}[t]
  \centering
  \includegraphics[width=1.0\linewidth]{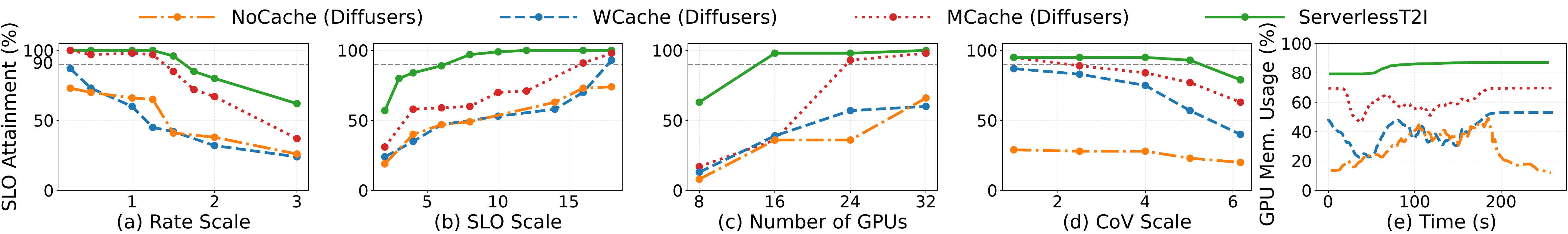}
  \caption{End-to-end evaluation of Setting 2 (\textbf{S2}). All baselines are implemented on top of Diffusers~\cite{diffusers}.}
  \Description{}
  \label{fig:s2_end2end}
  \vspace{-.2in}
\end{figure*}

As \figref{fig:s1_end2end} and \figref{fig:s2_end2end} show, \sys{} consistently achieves higher SLO attainment than the baselines in \textbf{S1} and \textbf{S2}. 
We use a controlled evaluation methodology: each experiment varies one workload or deployment factor while holding the others fixed. 
Overall, \sys{} sustains up to 2$\times$ higher request rate, satisfying up to 7$\times$ stringent SLOs, saving up to $3\times$ GPU resources, and tolerating 2$\times$ higher burst traffic, all while maintaining SLO attainment for 90\% of the requests.
These gains are partly driven by \sys{}'s model-loading design: data-plane weight caching accelerates 99\%/86\% of model loads in \textbf{S1}/\textbf{S2}, leaving only 1\%/14\% to require full model loading.
Among the baselines, \mCache and \wCache consistently outperform \noCache, confirming \sys{}'s insight that fine-grained use of idle GPU memory improves serving efficiency.

\PHB{SLO Attainment vs. Rate Scale.}
We first vary the request rate scale while fixing the SLO scale, testbed size, and traffic burstiness as the default values.
\figref{fig:s1_end2end}(a) and \figref{fig:s2_end2end}(a) show that \sys{} preserves high SLO attainment over a wider load range than all baselines. 
In \textbf{S1}, \sys{} maintains 100\% attainment up to a rate scale of 1.5. 
By contrast, the strongest baseline, \mCache, falls below 90\% once the rate scale reaches 1.0, and \noCache drops to only 10\% at rate scale 2.0. 
In \textbf{S2}, \sys{} again remains above 90\% through rate scale 1.5, whereas \mCache reaches 85\% at the same load and the other baselines are below 50\%. 
These results show that whole-workflow caching alone is insufficient under increasing load: even when some requests hit in cache, monolithic deployments still pay high loading and scaling costs when the active working set changes. 
\sys{} avoids this cliff by decomposing workflows and reusing GPU-resident model state across requests.

\PHB{SLO attainment vs. SLO Scale.}
We next vary the SLO scales while fixing the rate scale and testbed size, using the scaled original production trace. 
\figref{fig:s1_end2end}(b) shows that \sys{} meets substantially tighter SLOs in \textbf{S1}: at an SLO scale of 3.0, \sys{} completes all requests within deadline, whereas \wCache and \mCache require scale 24.0 to exceed 90\% attainment, and \noCache never reaches 90\% even at the loosest SLO. 
This gap reveals that the baselines are not merely short of compute capacity; their tail latency is dominated by model and workflow loading overheads, which only very loose deadlines can hide.
The same trend holds in \textbf{S2} (\figref{fig:s2_end2end}(b)): \sys{} achieves 97\% attainment at SLO scale 8.0, while \mCache and \wCache require scales of 16.0 and 18.0, respectively.
Even as adapters increase workflow diversity, \sys{} converts deadline slack into SLO attainment more efficiently than baselines.

\PHB{SLO Attainment vs. Testbed Size.}
We next vary the number of GPUs while keeping the workload fixed. 
In \figref{fig:s1_end2end}(c), \sys{} meets the 90\% SLO attainment target with only 8 GPUs; the strongest vLLM-Omni baseline requires 24 GPUs to match this, \mCache requires 32, and \noCache remains below 90\% even at 32 GPUs. 
In \textbf{S2}, \sys{} reaches 98\% attainment with 16 GPUs, whereas the strongest \diffusers{} baseline requires 24 GPUs to exceed 90\% and other baselines remain far below target at 32 GPUs. 
These results show that \sys{}'s gains extend beyond latency: scaling individual models rather than entire workflows reduces over-provisioning and allows the cluster to operate as a shared GPU pool.

\PHB{SLO Attainment vs. CoV.}
Finally, we evaluate robustness to bursty traffic by varying the coefficient of variation (CoV) of request arrivals while fixing the average rate scale, SLO scale, and testbed size. 
Following prior works~\cite{alpaserve, gujarati2020Clockwork}, we partition the original trace into time windows, fit arrivals to a Gamma process, and resample at scaled CoV values to control burstiness.
Higher CoV increases short-term queue buildup, stressing autoscaling and cache replacement. 
As shown in \figref{fig:s1_end2end}(d), \sys{} maintains at least 96\% attainment up to CoV scale 4.0, while \mCache falls below 90\% between scales 2.4–4.0 and \wCache is below 90\% at the lowest CoV. 
In \textbf{S2}, \sys{} stays above 90\% through CoV scale 5.0, whereas \mCache drops to 89\% at scale 2.5. 
The comparison between \textbf{S1} and \textbf{S2} indicates that adapter-heavy workloads make burst handling more sensitive to cache granularity: 
workflow-level caching cannot react quickly when bursts shift demand across variants, while \sys{} absorbs these shifts by reusing shared model components.

\PHB{GPU Memory Utilization.}
Figures~\ref{fig:s1_end2end}(e) and \ref{fig:s2_end2end}(e) show average GPU memory utilization at runtime. 
\sys{} consistently achieves the highest utilization, followed by \mCache, \wCache, and \noCache. 
This ordering reflects their caching granularities: \sys{} caches at the granularity of model layers, \mCache at models, and \wCache at entire workflows. 
Finer granularity reduces internal fragmentation by better utilizing residual memory.
In \textbf{S1} at rate scale 1.5, \noCache, \wCache, \mCache, and \sys{} achieve 13\%, 29\%, 71\%, and 88\% utilization, respectively. 
In \textbf{S2} at rate scale 1.75, they achieve 30\%, 41\%, 62\%, and 84\%, respectively.

\subsection{Data Plane}
\label{sec:microbenchmark}

\begin{figure}[t]
  \centering
  \includegraphics[width=0.495\linewidth]{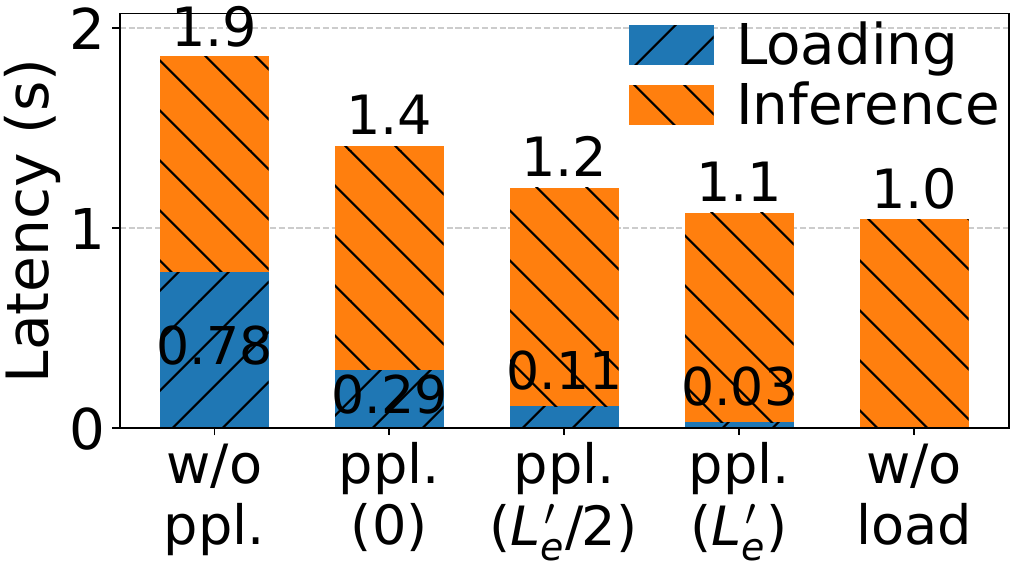}
  \includegraphics[width=0.495\linewidth]{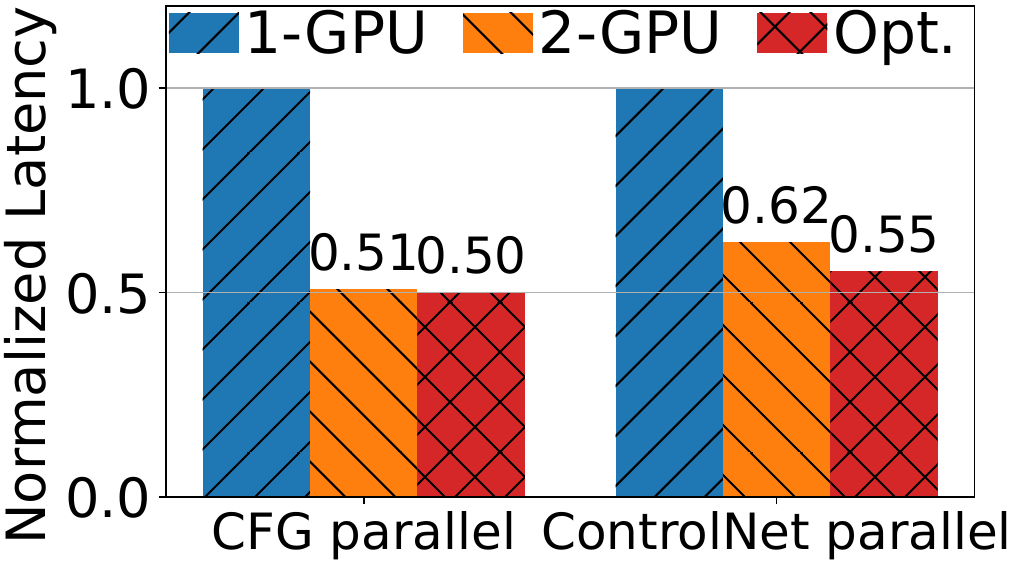}
  \caption{\textbf{Left:} Latency breakdown of workflow execution. \textbf{ppl.}: pipeline. \textbf{Right:} Normalized inference latency w/ and w/o parallelization. \textbf{Opt.}: theoretically optimal latency.}
  \Description{}
  \label{fig:micro_loading}
  \vspace{-.2in}
\end{figure} 

\PHB{Model Loading.}
We elaborate on the model loading design in \S\ref{sec:data_plane_loading}.
\figref{fig:micro_loading}-left reports the latency breakdown of a basic Flux1-Schnell workflow request, including model loading and inference.
We define two performance bounds: a lower bound where loading and inference execute serially without pipelining (\emph{w/o ppl.}), and an upper bound where model loading latency is fully hidden (\emph{w/o load}).
We show a spectrum of caching configurations: \emph{ppl.(0)} indicates no layers are pre-cached in the GPU, equivalent to existing layer-wise pipelined loading~\cite{pipeswitch, torpor}.
By caching $L^{\prime}_e$ layers in the GPU (\emph{ppl.($L^{\prime}_e$)}), \sys{} reduces loading latency by 90\% and end-to-end latency by 21\% relative to existing methods, rivaling the upper bound.
Even when half of the $L^{\prime}_e$ layers are evicted (\emph{ppl.($L^{\prime}_e/2$)}), \sys{} still reduces loading latency by 62\%.

\PHB{Communication.}
\sys{}'s data plane enables efficient and flexible communication for T2I-specific parallelization (\S\ref{sec:current_practice}). 
We validate these properties by measuring the speedups achieved by \sys{}'s parallelization, which would be substantially reduced if either property were absent. 
As shown in \figref{fig:micro_loading}-right, \sys{} exploits parallelism to accelerate Z-Image workflow executions on NVIDIA H800 GPUs. 
Its speedup approaches the theoretically optimized performance and is consistent with prior results~\cite{katz, li2024DistriFusion, fang2024xdit}, indicating that \sys{}'s data plane supports efficient, timely data exchange.

\begin{figure}[t]
  \centering
  \includegraphics[width=1.0\linewidth]{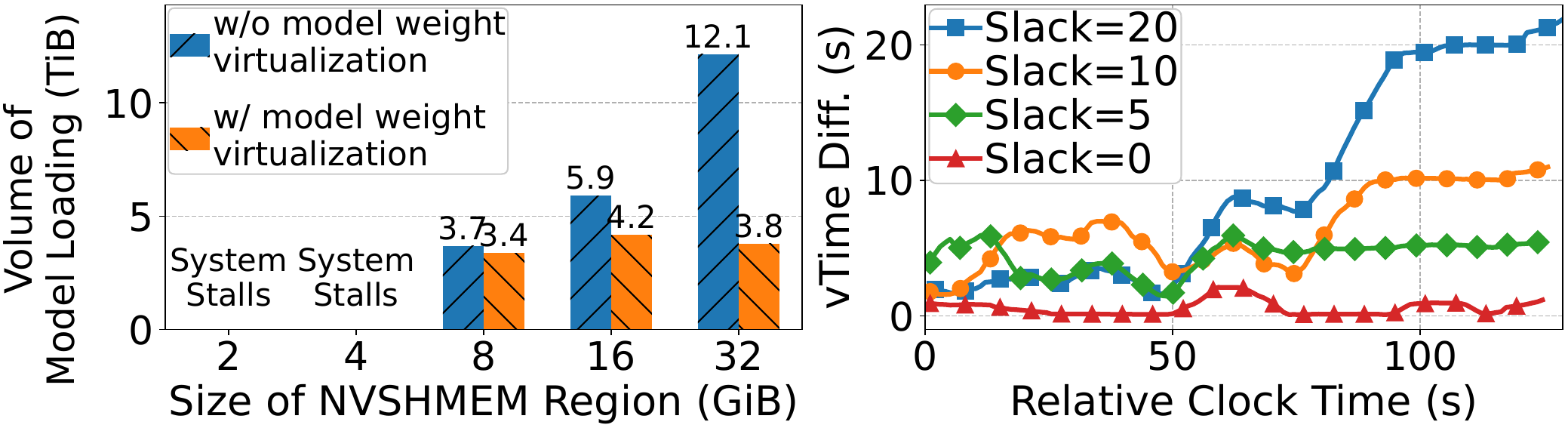}
  \caption{\textbf{Left:} Effectiveness of model weight virtualization.  \textbf{Right:} Fair scheduler with varying slacks.  }
  \Description{}
  \label{fig:weight_load_vruntime_diff}
  \vspace{-.2in}
\end{figure}

\PHB{Model weight virtualization.}
As described in \S\ref{sec:data_plane_unified}, \emph{model weight virtualization} allows cached weights to reside in either GPU memory region, addressing the limitations of a statically sized NVSHMEM region.
We validate this on an 8-GPU testbed using the ControlNet-augmented workflows in \textbf{S2} at a rate scale of 0.5.
As shown in \figref{fig:weight_load_vruntime_diff}-left, an undersized NVSHMEM region can stall the system when in-flight operations cannot allocate space for data communication, even after all weights have been evicted from the NVSHMEM region.
An oversized NVSHMEM region, conversely, crowds out the space for cached model weights without \emph{model weight virtualization}, increasing the volume of loaded weight by up to 3$\times$ and degrading SLO attainment by up to 50\%.

We also verified that repeated offloading and reloading of parameters does not fragment the PyTorch or NVSHMEM regions: even at 97\% peak memory utilization, we observed zero allocation stalls, retries, or out-of-memory events.

\subsection{Tradeoff between Fairness and Efficiency}
\label{sec:fair_scheduler}

\sys{}'s scheduler exposes a configurable fairness slack to balance fairness and efficiency. 
We set a microbenchmark on an 8-GPU testbed with two tenants, issuing basic Z-Image-Turbo and Flux1-Schnell workflows, respectively. 
In \figref{fig:weight_load_vruntime_diff}-right, \sys{} bounds the service difference  according to the configured slack when requests are backlogged. 
Larger slack values permit greater transient imbalance, but improve scheduling flexibility and hence serving efficiency: SLO attainment increases from 62\% with strict fairness (Slack=0) to 75\%, 81\%, and 84\% with slack values of 5, 10, and 20, respectively.

\subsection{System Overhead}
\label{sec:system_overhead}

\PHB{Execution Overhead.} Decomposing a monolithic workflow into a model DAG introduces overhead from data communication and control-plane coordination.
We quantify this overhead by comparing \sys{} with monolithic baselines on \textbf{S1}'s workflows.
For each workflow, we measure the execution latency of requests that generate the identical image. 
Compared with \diffusers{}, on which \sys{} builds its model components, \sys{} adds only 3\% average latency overhead. 
While \vllm{} reduces execution latency by 8\% relative to \sys{}, it falls short in the end-to-end evaluation (\S\ref{sec:e2e_performance}).

\PHB{\sys{} at scale.}
To show \sys{} remains
efficient at large scale, we conduct simulation-based experiments on a 256-GPU
setup under high concurrency, with 600 inflight requests. 
The simulator models \sys{}'s procedures with request latencies matching measured values.
Across the workflows in \textbf{S1}, \sys{} incurs
only 3.3\% overhead of total execution time, 
indicating that neither the control plane nor the data plane becomes a bottleneck at this scale.

%% file: contents/7_related_works.tex
\section{Discussion and Related Work}

\PHB{Can \sys{} serve video generation models?}
While video generation also uses diffusion models, its serving characteristics differ substantially from T2I, placing it outside \sys{}'s scope.
\emph{First}, video models exhibit different loading and computation profiles.
For the transformer of Wan2.2-14B~\cite{wanxiang}, the one-step denoising inference latency for a 480P video on an H800 is 2$\times$ higher than its loading latency, so loading can be pipelined with computation.
\emph{Second}, video generation service is typically exposed through APIs rather than user-composed workflows, a poor fit for serverless execution. As evidence, Wan2.2-14B (Mar. 2025) has only 2 community adapters on HuggingFace, whereas Z-Image (Nov. 2025) has 135, both released by the same Alibaba Tongyi team.

\PHB{Model Serving in Serverless Clouds.}
To our knowledge, prior serverless model-serving systems target DNN and LLM
inference rather than T2I workflows~\cite{deepserve}. 
To reduce model loading latency, 
PipeSwitch and Torpor~\cite{pipeswitch, torpor} overlap host-to-GPU model loading with inference, but this is
less effective for T2I workflows, where loading can still dominate the execution
time of individual stages. BlitzScale and FaaScale~\cite{zhang2024fast, faasscale} speed up loading by
transferring parameters over high-speed GPU interconnects instead of PCIe.
However, this design assumes a small set of repeatedly loaded model types and
spare interconnect bandwidth, both of which are less suitable for serverless
T2I serving, where workflows use diverse models and interconnects are also
needed for intermediate tensor communication. ServerlessLLM~\cite{serverlessLLM}
reduces LLM loading latency with a multi-tier path across disk, host memory, and
GPU memory. As discussed in \S\ref{sec:unified_data_plane}, \sys{} instead
focuses on memory-to-GPU loading and GPU-resident communication, where pinned
memory alone is insufficient. Its disk-to-memory optimization is complementary
to \sys{}. ServerlessLLM~\cite{serverlessLLM}, DeepServe~\cite{deepserve}, and
Medusa~\cite{Medusa} also optimize LLM-specific state such as KV-cache
management and CUDA-graph materialization. These techniques are valuable, but
they do not address the workflow heterogeneity and data-movement bottlenecks of
serverless T2I serving.

\PHB{T2I Workflow Serving Systems.}
Existing T2I workflow serving
systems~\cite{diffusers_server, comfyUI_server, sglang_diffusion, vllm_omni}
 accelerate individual workflow execution, but they do not target serverless deployment. As a
result, they inherit the limitations of deploying T2I workflows as monolithic
GPU functions on serverless platforms (\S\ref{sec:limitations}). 
Nirvana~\cite{nirvana}
reduces denoising steps through cached images; DistriFusion~\cite{li2024DistriFusion}
and xDiT~\cite{fang2024xdit} exploit multi-GPU parallelism; Katz~\cite{katz}
parallelizes ControlNets and asynchronously loads LoRAs; TetriServe~\cite{lu2026tetriserveefficientditserving}
and TridentServe~\cite{xia2025tridentservestagelevelservingdiffusion} adapt
sequence parallelism for latency SLOs. However, several of these
systems~\cite{xia2025tridentservestagelevelservingdiffusion,
lu2026tetriserveefficientditserving, nirvana, li2024DistriFusion} do not
support the adapters commonly used in production
workloads~\cite{katz, diffusion_production}. \sys{} is complementary to these
systems: it targets efficient T2I workflow serving on serverless platforms and
focuses on serverless-specific challenges in scaling, data movement, and
multi-tenant scheduling.

\PHB{Other Model Serving Systems.} Prior work on model serving has improved
latency~\cite{crankshaw2017Clipper, wang2023Tabi, she2026plaserve, schroeder2026vcache, roller,RAGO,zhang2025dissectingimpactmobiledvfs,fusion_RAG_Cache},
throughput~\cite{ahmad2024Proteus, yang2022INFless, Libra_nsdi, stateful_llm}, and resource
efficiency~\cite{zhang2019MArk, wang2021Morphling, gunasekaran2022Cocktail,
yang2025Prism, wang2023MGG} across DNNs and LLMs~\cite{yu2022Orca, sarathi,
MuxServe, loongserve,helix,HedraRAG,CacheBlend,oliaro2025FlexLLM,
JITServe,ic_cache, LLMStation, DeltaZip, ktransformer, srivatsa2025preble,
Mobile_Soc_llm, WEAVER, HyperGen}. \sys{} complements this line of work by focusing on
T2I workflow serving, which has different computation characteristics. KunServe~\cite{Cheng2024KunServePM} is the closest to our data
plane design, but it is specific to LLM serving: it observes redundant LLM parameters and selectively drops them to free
memory for KV cache.

%% file: contents/8_conclusion.tex
\section{Conclusions}
We presented \sys{}, an efficient serverless inference system for T2I workflows. 
\sys{} has three key designs: (1) a programming interface that allows users to compose workflows as a model DAG; (2) a unified data plane that harvests slack GPU memory for efficient model loading and data communication; and (3) a fairness-aware scheduler for multi-tenant serverless serving. 
Overall, \sys{} substantially improves T2I workflow serving efficiency. Under the same GPU budget, it sustains up to 
2$\times$ higher request rates than existing serving systems; at a fixed request rate, it reduces GPU requirements by up to 
3$\times$ while meeting SLOs for more than 90\% of requests.

%% file: contents/9_appendix.tex
\clearpage
\appendix

\section{Caching Workflows in Host Memory}

\label{sec:host_memory_loading}

This appendix provides additional details on the trace replay described in
\S\ref{sec:data_plane_loading}. 
A T2I workflow is large, often totaling tens of
GiBs, and loading its weights from external storage can introduce substantial
latency. 
However, production workflow popularity is highly skewed: a small
number of workflows account for the vast majority of requests. 
This skew makes
host memory an effective intermediate cache tier. By storing the weights of only
the most popular workflows in otherwise underutilized host memory, \sys{} can
avoid reading external storage for nearly all model loads, leaving only the much
faster host-to-GPU transfer on the critical path.

To quantify this effect, we replay our production trace using host memory caches
of varying capacities. 
We use least-recently-used (LRU) replacement and measure
the cache miss rate, defined as the fraction of requests whose workflow weights
are not present in host memory and therefore must be fetched from remote
storage. 
The miss rate drops quickly as cache capacity increases and then
saturates near zero: caching the $10$ most popular workflows reduces the miss
rate to $0.37\%$, while caching the top $15$ reduces it further to $0.09\%$.
Thus, with only the top $15$ workflows cached in host memory, more than $99\%$
of model loads are served from host memory rather than remote storage.

We also evaluate a least-frequently-used (LFU) policy and observe similar
results, suggesting that the benefit comes primarily from the inherent popularity
skew rather than from a specific eviction policy. Since host memory is largely
underutilized in our production setting (\figref{fig:resource_util}), caching
these popular workflows imposes little additional resource cost. Based on this
observation, \sys{} keeps popular workflow weights in host memory and focuses
the remainder of the loading design on optimizing the host-to-GPU transfer path.